\documentclass[sigconf]{acmart}

\AtBeginDocument{%
  \providecommand\BibTeX{{%
    \normalfont B\kern-0.5em{\scshape i\kern-0.25em b}\kern-0.8em\TeX}}}

\copyrightyear{2021} 
\acmYear{2021} 
\setcopyright{rightsretained} 
\acmConference[CHI PLAY '21]{Extended Abstracts of the 2021 Annual Symposium on Computer-Human Interaction in Play}{October 18--21, 2021}{Virtual Event, Austria}
\acmBooktitle{Extended Abstracts of the 2021 Annual Symposium on Computer-Human Interaction in Play (CHI PLAY '21), October 18--21, 2021, Virtual Event, Austria}\acmDOI{10.1145/3450337.3483507}
\acmISBN{978-1-4503-8356-1/21/10}

\begin{document}

\title{RelicVR: A Virtual Reality Game for Active Exploration of Archaeological Relics}


\author{Yilin Liu}
\affiliation{%
  \institution{Xi'an Jiaotong-Liverpool University}
  \city{Suzhou}
  \country{China}}
\email{yilinli@seas.upenn.edu}

\author{Yiming Lin}
\affiliation{%
  \institution{Xi'an Jiaotong-Liverpool University}
  \city{Suzhou}
  \country{China}}
\email{jameslimers@gmail.com}

\author{Rongkai Shi}
\affiliation{%
  \institution{Xi'an Jiaotong-Liverpool University}
  \city{Suzhou}
  \country{China}}
\email{rongkai.shi19@student.xjtlu.edu.cn}

\author{Yiming Luo}
\affiliation{%
  \institution{Xi'an Jiaotong-Liverpool University}
  \city{Suzhou}
  \country{China}}
\email{yiming.luo19@student.xjtlu.edu.cn}

\author{Hai-Ning Liang}
\authornote{Corresponding author: haining.liang@xjtlu.edu.cn}
\affiliation{%
  \institution{Xi'an Jiaotong-Liverpool University}
  \city{Suzhou}
  \country{China}}
\email{haining.liang@xjtlu.edu.cn}


\begin{abstract}
Digitalization is changing how people visit museums and explore the artifacts they house. Museums, as important educational venues outside classrooms, need to actively explore the application of digital interactive media, including games that can balance entertainment and knowledge acquisition. In this paper, we introduce RelicVR, a virtual reality (VR) game that encourages players to discover artifacts through physical interaction in a game-based approach. Players need to unearth artifacts hidden in a clod enclosure by using available tools and physical movements. The game relies on the dynamic voxel deformation technique to allow players to chip away earth covering the artifacts. We added uncertainty in the exploration process to bring it closer to how archaeological discovery happens in real life. Players do not know the shape or features of the hidden artifact and have to take away the earth gradually but strategically without hitting the artifact itself. From playtesting sessions with eight participants, we found that the uncertainty elements are conducive to their engagement and exploration experience. Overall, RelicVR is an innovative game that can improve players' learning motivation and outcomes of ancient artifacts.
\end{abstract}



\begin{teaserfigure}
  \includegraphics[width=\textwidth]{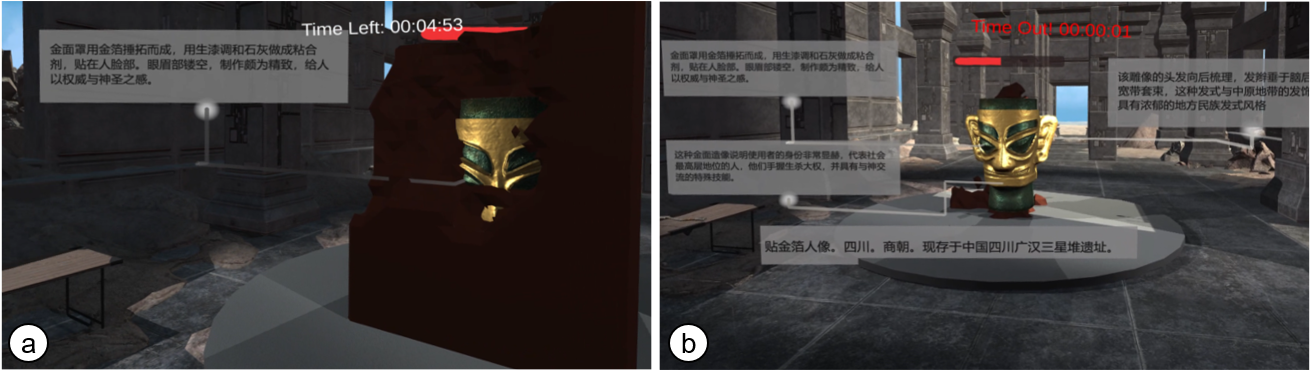}
  \caption{Screenshots of RelicVR: (a) the bronze head with a gold mask is partially revealed with a dialog box shown to introduce it, (b) the relic is completely excavated and all four dialog boxes describing the relic are shown.}
  \Description{The two sub-figures show two phases in the game process. Figure labelled 'a' shows an early stage of the game. The player has excavated the relic and triggered a key point, so a dialog box is shown to describe some features of the relic. Figure labelled 'b' is a view of the game that is near to completion. Most of the clod enveloping the artifact has been cleaned. All four dialog boxes describing the relic are displayed.}
  \label{fig:teaser}
\end{teaserfigure}

\maketitle

\section{Introduction}
Museums, as one of the most important educational institutions outside schools, aim to exhibit and introduce historical and cultural artifacts from multi-sectors to the general public. However, in a time when most students are visual learners and are familiar with games, they are more attracted by and engaged in the content with active exploration and interaction \cite{Jukes01}. The young generation has gradually become impatient with traditional modes of instructions, such as using text or audio explanations, and disengaged with these conventional paradigms of exploration and knowledge acquisition \cite{eck06}. Today's museums, like other learning institutions, are trying to keep with technological and social changes and looking into ways to transform inactive exhibitions into interactive experiences to better meet their audiences' preferences, especially those of younger generations. Game-based learning is one of the interactive approaches that can satisfy their needs. A well-designed game-based learning context would balance gaming and learning and can benefit players twofold by (1) motivating them to learn the knowledge with a high level of engagement with the enjoyable game content \cite{Prensky01}, and (2) guiding them with a deeper understanding of complex settings in the games \cite{Gros01}. These advantages of game-based learning make games suitable to be embedded in the context of museums to make the exploration of artifacts active and engaging.

In the discovery of ancient artifacts, uncertainty is always a challenge for archaeologists. Excavation or unearthing of relics is one of the most important and exciting parts to eliminate uncertainty. Based on this feature, Henan Museum has launched "the archaeological blind box," a small brick with a random mini-artifact hidden inside it. It requires the visitors to use mini tools to sweep the dirt away to discover and collect the hidden artifact. By offering the young generation opportunities to stimulate them to explore the relics and implicitly encourage them to learn cultural knowledge, the blind boxes started a wave of amateur archaeology in China with 170,000 pieces sold from December 2020 to April 2021 \cite{labbrand01, Xinhua01}. On the other hand, uncertainty is also an essential element in video games. Well-designed uncertainty elements in a game could motivate players to start and continuously keep players engaged in the game \cite{Caillois01, Costikyan15, Johnson18, Power19}. The combination of archaeology and gaming provides a chance to blend knowledge and uncertainty together in a game to motivate players to acquire knowledge and enjoy the learning process.  

Virtual reality (VR) games can be an ideal medium to achieve such a combination and enhance understanding of cultural knowledge through an immersive, active, and fun experience. VR expands the space and affordances of real-world museums and transforms them into immersive virtual environments that can incorporate creative and imaginative elements. 3D virtual gaming worlds can provide unlimited interactive possibilities and experiences at a low cost. People can access these experiences remotely, which is valuable when there are restrictions on time, cost, or distance, especially during the COVID-19 period. With its active experimentation and high immersion, VR games can enhance constructive, situated, and experimental learning \cite{Silva2017}. 

We built a VR archaeology educational game, RelicVR, to enrich the museum visiting and learning experience. Our game provides a game-based learning environment, which emulates a simplified archaeological process of discovering ancient artifacts but with innovative gameplay elements. The archaeological process involves a set of actions such as chipping and digging away the earth that envelops a hidden artifact (see Figure \ref{fig:teaser}). The players need to wield tools in their hands and move around to chip away the clod of earth that wraps the relic. Since players initially have no knowledge of the artifact inside the clod, the uncertainty during gameplay is intended to heighten their senses and focus on the chipping process to prevent damaging the artifact. In short, RelicVR aims to make the exploration of ancient artifacts an active, fun, and engaging process. We conducted a short playtesting with eight players to evaluate the effects of gaming for learning about relics. Results showed that the game can provide a fun and engaging experience to players and is still be able to help them learn the features of the artifacts.

\section{Related Work}
\subsection{Games for Archaeology}
Developments in the field of archaeology have changed the way we present archaeological artifacts and discoveries today. The growing interest in digitalization and turning the process interactive has a considerable appeal to archaeologists wishing to present their research to the public. Early in 2002, Watrall \cite{Watrall02} stated that archaeologists must realize the potency of interactive media and embrace it. The educational potential of interactive games has been increasingly addressed over the past years, especially for pre-teen learners, as game-based learning can keep them more engaged and make their efficient learning \cite{eck06}. Some prior research has shown the overall success of applying educational games to enable students to learn about the archaeological investigation (e.g. \cite{Ardito12,Georgiadi}). At the University of Illinois, the VRchaelogy Lab \cite{ShacHuanCrai2018ra, ShacHuanCrai2019, Ramy16} had some surprising findings of how VR environments can enhance the learning of basic archaeology knowledge to college students and non-professionals. 

\subsection{Games for Museums}
Efficient approaches that balance educational purposes and entertainment are desired in today's museums. The use of video games can increase general interests, especially from young generations, in exploring cultural artifacts and improve their experience \cite{Varinlioglu19}. Studies have indicated that a gamified application can increase participants' involvement, motivation, enjoyment, and overall exploratory experiences \cite{Brigham15, Seaborn15}. 

Recent advances in VR have led to new possibilities in museum touring because of its capability to create immersive artificial worlds. There have been projects in which VR is used to recreate virtual living spaces in museums \cite{George03, Viking, BackToLife}. Jung et al. \cite{Jung01} found that entertainment experienced from VR can enhance the overall museum visits. However, most of the existing games for archaeology lack engaging interaction between players and the virtual environments. There are nearly no games that introduce artifacts from an archaeologist's discovery perspective. We argue that the scenery in VR alone is not enough to engage players and arouse their interests in the cultural heritage. In \textit{The Grasshopper: Games, Life, and Utopia Suits} \cite{cooper82}, Cooper defines the playing of a game as “the voluntary attempt to overcome unnecessary obstacles." To lead players to this self-motivated act, it is also necessary to include appropriate methods in the design of a successful game experience. Compared with existing games (e.g., \cite{ShacHuanCrai2018ra, ShacHuanCrai2019, Ramy16}), which are somewhat more oriented to teach more serious concepts and emphasize more on the learning side than on gameplay, RelicVR has more gaming features and is intended to be more interactive, encourage exploratory engagement and focus, and allow learning to be fun. 

\subsection{Uncertainty}
Uncertain elements can make the gameplay experience engaging, enjoyable, and playful \cite{Caillois01, Costikyan15, Kumari2019}. According to Costikyan \cite{Costikyan15}, typical uncertain sources in a game include performative uncertainty, solver’s uncertainty, player unpredictability, and randomness. In terms of VR content, Xu et al. \cite{xu21} evaluated the effect of uncertainty in a VR exergame. They found that uncertainty could improve the exergame experience by increasing players’ energy exertion. We believe involving uncertainty elements in our game would give players a better gaming and learning experience. To achieve this effect, we obscure the object with a clod of earth that wraps the object entirely. Since the players do not know the structure and features of the hidden object, they need to carefully plan their actions and develop reasonable speculations based on the parts that have been revealed to avoid hitting the relic and damage it. In addition, if the object is hit, the game provides a dissonant sound, akin to negative feedback---something that players would want to avoid. According to Costikyan's classification, our approach could be classified as performative uncertainty because there is uncertainty in the outcome of the physical actions of the players and the gradual unearthing of the artifact. As the players reveal more hidden parts of the artifact, they are also going through a process of verifying their hypotheses or speculations in an iterative manner and continuously update their knowledge of the artifact.

\section{RelicVR}
We designed and developed RelicVR, a first-person perspective VR game that combines active exploration and knowledge acquisition of archaeological artifacts. The game intends to enrich the learning experience with such artifacts by embedding them into a game that simulates a simplified procedure of archaeological discovery. RelicVR is made to be run in the Oculus Rift S. It was developed using C\# on the Unity3D platform (version 2020.3.5f1) with the Oculus Integration SDK (28.0.0). In this section, we introduce its overall concepts, game world scenario, and interactions in the virtual environment (VE). 

\subsection{Concept}
The principle behind RelicVR was inspired by archaeological activities in the real world. Instead of implementing the whole archaeological investigation approach, we mainly adapted the excavation phase in the game due to its interactive nature. In RelicVR, players start with little information of what is hidden in the clod of earth. When they use tools to remove the chunks of earth from the clod, they face a sense of uncertainty due to the unknown shape, structure, and features of the relic. As such, they need to be focused on the actions and observe the outcome of each physical activity. Such a process can be iteratively reinforced as more parts of the relic come to light.

We set up a few invisible trigger points in different places around the relic according to its features. When players hit a trigger point, a dialog box with text describing some features would appear and is accompanied with audio to capture players' attention. As more and more details of the artifact are exposed, players gradually increase their understanding by reading or listening to the prompted descriptions and careful observations of its visual details, now visible to the naked eye. In this first version of the game, we designed three dialog boxes for each relic to introduce it from different perspectives. 

\begin{figure*}
    \centering
    \includegraphics[width=\linewidth]{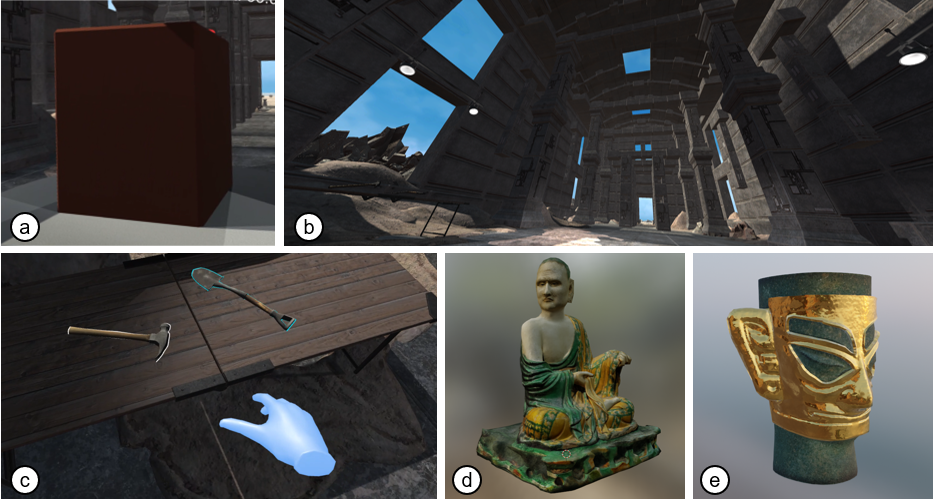}
    \caption{Game elements in RelicVR. (a) Players will see a clod at the beginning of the game. (b) Game environment is a mysterious temple to further immersive players in the activity. (c) Two available tools (a hammer and a shovel) to be used to excavate the artifact. (d) and (e) are two relics included in the first version of RelicVR. They are the Arhat (Luohan) sculpture and Sanxingdui bronze head with a gold mask, respectively.}
    \Description{This figure provides some screenshots of the first version of RelicVR including the clod that hides the artifact, the game environment, digging tools, and two sample relics. }
    \label{fig:relicvr}
\end{figure*}

RelicVR aims to drive players to engage in active exploration and learning. It gives the players a similar role to an archaeologist to explore the relic actively. The players start out by observing a large clod of earth placed at the center of the VE, but can only have a limited understanding of the relic (see Figure~\ref{fig:relicvr}a). Driven by curiosity, the players are encouraged to discover what lies inside the clod. The game requires physical motions in the form of digging with selected tools and walking around to try to unearth the artifact within a limited time. This process requires continuous planning and problem-solving and frequent updates to the mental model of the object. During the exploration process, they also face similar challenges like archaeologists---they need to carefully plan their actions to avoid damaging the artifacts. After players have successfully excavated the artifact, they can get general information from an additional dialog box and put the relic into their collection. While not implemented in the current version, the collection can be shown to their friends, which can give them a sense of achievement.

\subsection{Game World}
In the first version of the game, we used a mysterious temple (Figure~\ref{fig:relicvr}b) as the gaming environment to increase the adaptability of artifacts from different eras and cultures. The temple does not refer to any cognizable mono-cultural motif. The daylight in the scene moves around the temple to give players a sense of time elapsing as the game progresses. Several spotlights for illuminating the artifact are hung high in the temple to focus players' attention. Fallen walls and rocks are randomly placed outside the temple to give a sense of ancientness and historical allure and mystery to the environment. As the players look around, they can see that the scene is a semi-enclosed space containing tools on a table and an unexplored artifact encased in a cubic clod of earth.  

Two relics are included in the current version of RelicVR. The first one is the Arhat (Luohan) Sculpture from the Metropolitan Museum of Art in New York (Figure~\ref{fig:relicvr}d, referred to as Arhat hereafter) \cite{Luohan}. The second relic is the bronze head with a gold mask from Sanxingdui Museum in China (Figure~\ref{fig:relicvr}e, referred to as Gold Mask hereafter) \cite{GoldenMask}. We extracted the information from their official website and integrated it into the game. These two relics are scaled to approximately the same volume to minimize the differences in the gaming experience. Their sizes are about a human's height to increase the level of exploration. In addition, it provides players a chance of detailed observations of the relic which is less likely to be achieved in conventional museum visits. 

The players can choose a hammer or a shovel on the workbench based on their preference, as shown in Figure~\ref{fig:relicvr}c. The hammer makes players feel more powerful and more flexible to handle, while the shovel is thinner but has a broad blade. The shovel can easily remove large areas of clod chunks but also increases the risk of hitting the artifacts. This is a trade-off between the two tools that the players need to balance carefully.

\subsection{Interactions}
The interactions in RelicVR mimic real-world actions to increase immersion and physical activity. Players need to grab one of the tools by pressing the grip button on the controller after pointing the virtual hand at the tool at a short distance or remotely at a longer distance. To simulate how digging is performed in real life, players are required to hold the grip button during the whole excavation activity. Players need to wave their arms to dig. The actions in the VE match the players' actions in the physical world. Since the artifacts and clod are designed to be of human size, players might need to raise their arms and squat to clean the dirt at the top and bottom regions. 

In the virtual world, the players are encouraged to move physically around the clod enclosure to achieve coordinated movements in the virtual scene. The players can also use the joysticks on the controllers to move or turn around. However, such movements or turnings are only suggested when the player needs to make large positional adjustments in the VE. 

Similar to real archaeological excavations, the players should control their actions carefully to avoid damaging the relic during the excavation process. Warnings are given using a (dissonant) sound that simulates striking the artifact. The choice of the negative sound feedback is to encourage players to avoid hearing it by not hitting the artifact. 

The levels of difficulty in RelicVR can vary greatly according to the complexity of shapes and volume. For the current version, we selected models of similar complexity and calculated their approximate volume. We set a health bar and timer for the relic as elements of the game: the reduction of the health bar and timer has no effect on the artifacts and gameplay but would create tension and motivation for players to complete the task within the allotted time and minimize the reduction of health points. The health bar hovers above the artifact and always faces the player's view. A strike on the relic would reduce 1 point. Time begins to count down from when the game starts and is shown on the upper middle part of the screen. We found that for the current models, 7 minutes of playtime and 40 health are suitable for most players to keep a balance between having a degree of pressure and feeling motivated to complete the task.

\subsection{Innovation}
Recent advancements in graphical computation power and development of the marching cubes algorithm \cite{lorensen87} allow the deformation of objects that are non-trivial and are very detailed in real-time. To our knowledge, RelicVR is the first game that explores the use of dynamic voxel functions in VR to combine active gaming and exploratory learning together. Furthermore, instead of exploring a vast area of a VE such as \cite{DeepRock, Harbin}, RelicVR attempts to focus players in  active explorations of a human-scale 3D area. The physical interaction afforded by VR devices enables physical activity by players during gameplay. Furthermore, the inclusion of uncertainty elements in the game can make players immersed and focused on playing the game.   

\section{Playtesting}
\subsection{Experiment}
We have conducted a preliminary playtest with eight players (3 females and 5 males, aged between 19 to 25; \textit{M} = 21.5, \textit{SD} = 2.07). They were recruited from a local university. Half of them experienced the Arhat version, while the remaining experienced the Gold Mask version. The players were first introduced to the game and the use of the devices.  They were asked to use tools to excavate the artifact enclosed within the clod. In addition, they were told to keep in mind the time they had and try to avoid damaging the artifact by not hitting it during the excavation process. At the same time, they needed to remember the details from the dialog boxes that would appear after discovering a key point, if possible. After experiencing the game, they were asked to complete a quiz with 6 fill-in-the-blank questions to assess how much they remembered after playing the game. Answers to these questions were derived from the content of the dialog boxes. In addition, we gave them two questionnaires, a core module of the Game Experience Questionnaire (GEQ) \cite{IJsselsteijn} and an adapted version of web-based learning tools evaluation scales (WBLT) \cite{Kay11} to elicit their perceived gaming and learning experience with RelicVR. At the end of the experiment, we conducted a semi-structured interview asking them about their feelings and thoughts about the game. 

\subsection{Results and Discussion}
\begin{figure*}
    \centering
    \includegraphics[width=\linewidth]{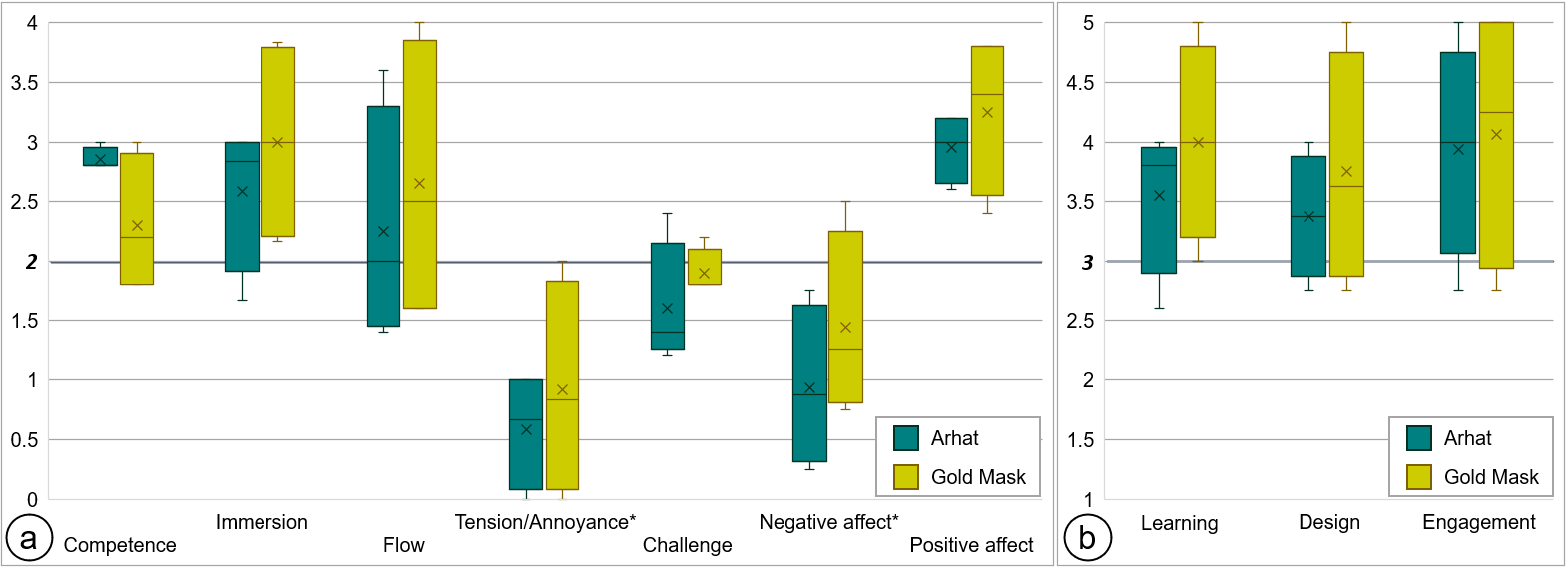}
    \caption{Boxplots of the questionnaire results. (a) GEQ on a scale of 0 to 4. (b) WBLT on a scale of 1 to 5. For both questionnaires, the higher score represents a better experience in terms of the specific component. Scales representing neutral feeling are highlighted in both sub-figures. Component with asterisk symbol represents a reverse rating.}
    \Description{Figure labelled 'a' shows the ratings of the 33-item core module of the GEQ. The ratings of the items were grouped into 7 categories: competence, immersion, flow, tension/annoyance, challenge, negative affect, and positive affect. Overall, RelicVR received positive feedback about game experience. Figure labelled 'b' shows the ratings of the 13-item WBLT evaluation scales. The ratings were grouped into 3 categories: learning, design and engagement. Most of the players gave high ratings to RelicVR.}
    \label{fig:questionnaire}
\end{figure*}

Overall, the eight players were satisfied with the experience in RelicVR. Figure~\ref{fig:questionnaire} summarizes the results of GEQ and WBLT. Based on the ratings, the players perceived a positive gaming experience in RelicVR in terms of competence, immersion, flow, and positive affect (the ratings in these categories were all above neutral), while they felt less tension/annoyance and negative affect (below neutral). On average, players gave lower-middle ratings in the challenge category, which shows a moderate level of difficulty in the game. Based on our results, RelicVR also provided a good learning experience. All eight players were able to answer most of the questions correctly (\textit{M} = 4.88, \textit{SD} = 0.83) after experiencing the game. As shown in Figure~\ref{fig:questionnaire}b, the players perceived RelicVR as a good learning tool in learning, design, and engagement (all above neutral).  

During the interview, all eight players reported that RelicVR was more interesting than conventional museum visits, where artifacts could only be seen from a distance and are enclosed behind glass windows, a process that is typically non-interactive and impersonal. On the other hand, four players (P2, P4, P5, P6) mentioned that in RelicVR, the process of excavating by hand forced them to be close to the relic and be focused on it. Coupled with the assistance of text and audio introduction, they reported a deep impression of the relic and its information. P4 said, ``\textit{The dialog boxes at different key points were good to let me know about the artifact. When they came out, I could have some rest and time to watch and appreciate their physical appearance, of which I still have a clear memory after the game.}'' P3 and P7 said that they enjoyed the uncertainty in the exploration process. Based on our observations, they were always looking forward to seeing what the artifact would look like and trying carefully not to hit the object when clearing the earth. 

P1 and P2 said that the game should make more explicit the connections between playing in the game and learning and have a better balance between them, though they thought RelicVR was a great learning tool. They mentioned that, by separating the description of the relics into multiple dialog boxes connected to the corresponding components of the relic, they got a better learning experience than reading a long description in one single place. However, they also argued that the learning experience should be further improved. Externally, because the timer was continuously counting down, they felt somewhat anxious to excavate the unfinished parts. Internally, they could not wait to explore more about the relic because the gaming elements were more appealing to them. To a certain extent, balancing the entertainment and education elements is often a challenge in games for learning. To provide a better learning experience, we plan to explore when the best time is to deliver knowledge related details in RelicVR; it should either be embedded in the game just like the current version, or be provided before or after the gaming part. On the other hand, it is also necessary to compare and derive the most suitable completion time which can give tension or pressure to players to complete the tasks but at the same not to let them feel overwhelmed or overly anxious. P4 expected more tools in the game, especially the tools that can be used to do elaborate and more precise excavations. In short, while improvements can be made to RelicVR, the players seem to have enjoyed the combination of active exploration and uncertainty of the discovery process that the game brings.

\section{Future Work}
For macro-level interaction, we plan to include the possibility of collecting the artifacts that players unearth as a first priority. As a shared value of professional archaeologists and Henan Museum's blind boxes show, the function will enhance the sense of achievement of players and could motivate them to play the game in the long-term, as the game could include a wide range of artifacts, some of which could be released for special occasions and in a limited way. Having more artifacts can also be used to provide levels in the game. These artifacts could be from different cultures of various sizes and shape complexities which would require players to perform more demanding activities and digging patterns to unearth them.

Moreover, a richer micro-level interaction is desirable. We plan to include more excavation tools and vary their effects on the clod to provide diverse experiences for players. In addition, to increase the level of difficulty and challenge, we can follow a similar approach to the different block types in Minecraft \cite{Minecraft}. We can apply various materials and mix them into the clod enclosure. Different mixes will lead players to use different strategies and elicit greater levels of exertion during gameplay since the required energy and feedback are different for each mix. On the other hand, a recent study has proposed an attachable device to the VR handheld controllers to provide weight and center of gravity simulation in VR \cite{monteiro2021icmi}. We think such device or the concept can be used in RelicVR to provide force feedback on users' hands or arms to simulate real excavation actions. 
 
From a research perspective, RelicVR can be an ideal testbed for evaluating the effect of unknown and uncertain factors on gameplay and learning experiences in VR. In addition, it can be used to explore the effect of exertion levels \cite{xu21,xu2020chiplay}, body positions \cite{xu2020jmir}, and viewing perspectives \cite{xu2020g4h,monteiro2020ismar}. In addition, research has looked into the effects of display types for exergames \cite{xu21, xu2020g4h}. It is also interesting to evaluate and compare the experience and outcomes of game-based learning between VR and traditional displays (i.e., immersive vs. less-immersive experiences).

Finally, we also plan to refine other aspects of the game like optimizing further of completion condition of each artifact, letting players customize completion time, and adding a health bar system to balance game challenge and gameplay experience.

On a business level, RelicVR has great potential to be used in cooperation with museums to enhance the visitors' experience and promote popular science education. We have plans to contact local museums to explore ways that RelicVR can be used in any of their processes and enhance their visitors' experiences.

\section{Conclusion}
In this paper, we presented RelicVR, the first VR game that encourages players to discover historical artifacts through physical interaction and a game-based approach. By using dynamic voxel deformation in the game, it was able to add uncertain elements in the exploratory process---e.g., the unearthing of the artifact in a gradual but calibrated way to avoid hitting and damaging it. Our playtesters found that the uncertainty in the process helped enhance their experience. Also, the innovative interaction method has the potential to improve players' learning motivation and outcomes. Overall, RelicVR is an innovative game that can engage and immerse players in active exploration of archaeological artifacts.  

\begin{acks}
The original inspiration of RelicVR came from a visit to Henan Museum invited by Mr. Haosen Zhao, a passionate young archaeologist. We want to express our thanks to him and wish him success in pursuing his dream to contribute to the field of archaeology. The authors also want to thank the participants who helped playtest our game and the reviewers for their insightful comments that have helped improved our paper. This work was supported in part by Xi'an Jiaotong-Liverpool University's Key Special Fund (\#KSF-A-03).
\end{acks}

\bibliographystyle{ACM-Reference-Format}
\bibliography{ref}










\end{document}